\documentclass[twocolumn]{aastex631}

\usepackage{silence} 
\usepackage{times}
\usepackage{enumitem}
\usepackage{graphicx}
\usepackage{amsmath}
\usepackage{amssymb}
\usepackage{bm}
\usepackage{graphics}
\usepackage{url}
\usepackage{epsfig}
\usepackage{wrapfig}
\usepackage{datetime}
\usepackage{ulem}
\usepackage{verbatim}
\usepackage{float}
\usepackage{lastpage}
\usepackage{enumitem}
\usepackage{natbib}
\usepackage{mathrsfs}
\usepackage[OT1]{fontenc}
\usepackage{lmodern}

\providecommand{\apjl}{ApJ}

\providecommand{\apjs}{ApJS}
\providecommand{\aap}{A\&A}

\def\simlt{\mathrel{\hbox{\rlap{\hbox{\lower4pt\hbox{$\sim$}}}\hbox{$<$}}}}
\def\simgt{\mathrel{\hbox{\rlap{\hbox{\lower4pt\hbox{$\sim$}}}\hbox{$>$}}}}

\def\hst{{\it HST}}

\def\I{\,\textsc{i}}
\def\II{\,\textsc{ii}}

\def\V{\,\textsc{v}}

\def\arcsec{$^{\,\prime\prime}$}

\def\arcdeg{\degr}  

\shorttitle{SN\,2025pht Progenitor System}
\shortauthors{Kilpatrick et al.}

\begin{document}

\title{The Type\,II SN\,2025pht in NGC\,1637: A Red Supergiant with Carbon-rich Circumstellar Dust as the First {\it JWST} Detection of a Supernova Progenitor Star}

\correspondingauthor{Charles~D.~Kilpatrick}
\email{ckilpatrick@northwestern.edu}

\def\ciera{Center for Interdisciplinary Exploration and Research in Astrophysics (CIERA), Northwestern University, Evanston, IL 60208, USA}
\def\northwestern{Department of Physics and Astronomy, Northwestern University, Evanston, IL 60201, USA}
\def\jhu{Department of Physics and Astronomy, The Johns Hopkins University, Baltimore, MD 21218, USA}
\def\caltech{Department of Astronomy and Astrophysics, California Institute of Technology, Pasadena, CA 91125, USA}
\def\psu{Department of Astronomy and Astrophysics, The Pennsylvania State University, Davey Lab, State College, Pa 16802, USA}
\def\ucb{Department of Astronomy and Astrophysics, University of California, Berkeley, CA 94720, USA}
\def\ucsc{Department of Astronomy and Astrophysics, University of California, Santa Cruz, CA 95064, USA}
\def\ifa{Institute for Astronomy, University of Hawaii, 2680 Woodlawn Drive, Honolulu, HI 96822, USA}
\def\dark{DARK, Niels Bohr Institute, University of Copenhagen, Jagtvej 128, 2200 Copenhagen, Denmark}
\def\oxford{Department of Physics, University of Oxford, Denys Wilkinson Building, Keble Road, Oxford OX1 3RH, UK}
\def\icds{Institute for Computational \& Data Sciences, The Pennsylvania State University, University Park, PA, USA}
\def\qub{Astrophysics Research Centre, School of Mathematics and Physics, Queen’s University Belfast, Belfast BT7 1NN, UK}
\def\igc{Institute for Gravitation and the Cosmos, The Pennsylvania State University, University Park, PA 16802, USA}
\def\gemini{Gemini Observatory, NSF’s NOIRLab, 670 N. A’ohoku Place, Hilo, HI 96720, USA}
\def\carnegie{The Observatories of the Carnegie Institution for Science, 813 Santa Barbara St., Pasadena, CA 91101, USA}
\def\uiuc{Department of Astronomy, University of Illinois at Urbana-Champaign, 1002 W. Green St., IL 61801, USA}
\def\cas{Center for Astrophysical Surveys, National Center for Supercomputing Applications, Urbana, IL, 61801, USA}
\def\cambridge{Institute of Astronomy and Kavli Institute for Cosmology, Madingley Road, Cambridge, CB3 0HA, UK}
\def\toronto{David A. Dunlap Department of Astronomy and Astrophysics, University of Toronto, 50 St. George Street, Toronto, Ontario, M5S 3H4 Canada}
\def\uwmilwaukee{Department of Physics, University of Wisconsin-Milwaukee, 3135 North Maryland Avenue, Milwaukee, WI 53211, USA}
\def\nsfai{The NSF Institute for Artificial Intelligence and Fundamental Interactions}
\def\cfa{Center for Astrophysics $|$ Harvard \& Smithsonian, Cambridge, MA 02138, USA}
\def\mit{Department of Physics and Kavli Institute for Astrophysics and Space Research, Massachusetts Institute of Technology, 77\\ Massachusetts Avenue, Cambridge, MA 02139, USA}

\author[0000-0002-5740-7747]{Charles~D.~Kilpatrick}
\affil{\ciera}

\author[0009-0005-8230-030X]{Aswin~Suresh}
\affil{\ciera}
\affil{\northwestern}

\author[0000-0002-5680-4660]{Kyle~W.~Davis}
\affil{\ucsc}

\author[0000-0001-7081-0082]{Maria~R.~Drout}
\affil{\toronto}

\author[0000-0002-2445-5275]{Ryan~J.~Foley}
\affil{\ucsc}

\author[0000-0003-4906-8447]{Alexander~Gagliano}
\affil{\nsfai}
\affil{\cfa}
\affil{\mit}

\author[0000-0002-3934-2644]{Wynn~V.~Jacobson-Gal\'{a}n}
\altaffiliation{NASA Hubble Fellow}
\affiliation{Cahill Center for Astrophysics, California Institute of Technology, MC 249-17, 1216 E California Boulevard, Pasadena, CA, 91125, USA}

\author[0009-0005-1871-7856]{Ravjit~Kaur}
\affil{\ucsc}

\author[0000-0002-5748-4558]{Kirsty~Taggart}
\affil{\ucsc}

\author[0000-0003-1576-0830	]{Jason~Vazquez}
\affil{\uwmilwaukee}

\begin{abstract}

We present follow-up imaging and spectroscopy and pre-explosion imaging of supernova (SN) 2025pht located in NGC\,1637 at 12\,Mpc.  Our spectroscopy shows that SN\,2025pht is a Type\,II SN with broad lines of hydrogen and with minimal line-of-sight extinction inferred from Na\I~D absorption.  NGC\,1637 was the target of several epochs of {\it Hubble Space Telescope} (\hst) and {\it James Webb Space Telescope} ({\it JWST}) imaging covering the site of SN\,2025pht from 31 to 0.7~yr prior to discovery.  Using a follow-up \hst\ image of SN\,2025pht aligned to these data, we demonstrate that there is a credible progenitor candidate detected in multiple epochs of \hst\ imaging and in {\it JWST} imaging from 1.3--8.7\,$\mu$m, the first {\it JWST} counterpart to a SN and the longest wavelength detection of a SN progenitor star.  Fitting this source to red supergiant (RSG) spectral-energy distributions (SEDs), we show that it is consistent with a $\log(L/L_{\odot})=5.0$ RSG heavily reddened by circumstellar dust.  The {\it JWST} photometry enable strong constraints on the nature of the circumstellar medium, and we find that the SED favors graphite-rich dust and an optical circumstellar extinction of $A_{V}=5.3$~mag as opposed to silicate-rich dust.  We discuss the implications of a pre-SN RSG enshrouded in carbon-rich dust and this finding for the overall population of progenitor stars to Type\,II SN as well as the future of SN progenitor star discovery with {\it JWST}.

\end{abstract}

\keywords{
  stars: evolution --- supernovae: general --- supernovae: individual (SN\,2025pht)
}

%%%%%%%%%%%%%%%%%%%%%%%%%%%%%%%%%%∂%%%%%%%%%%%%%%%%%%%%%%%%%%%%%%%%%%%%
\section{Introduction}\label{sec:introduction}

Since the discovery of the blue supergiant progenitor star Sanduleak $-$69 202 to supernova (SN) 1987A \citep{Kirshner1987,West1987,Woosley1987,arnett+89}, over 30 other progenitor stars to core-collapse SNe have been identified in pre-explosion imaging \citep[e.g.,][]{Smartt09,vanDyk12,Tomasella13,Maund13,Fraser13,Kilpatrick18:17eaw,ONeill19,Sollerman21,Kilpatrick23:20jfo,VanDyk23a,Kilpatrick23,Xiang24}.  The vast majority of these systems are red supergiant (RSG) progenitor stars to Type\,II SNe \citep[see reviews in, e.g.,][]{Smartt15,VanDyk17}, the predicted terminal state for single stars with initial masses $\approx$8--30~$M_{\odot}$ \citep[corresponding to RSGs with terminal luminosities of $\log(L/L_{\odot})\approx4.3$--5.6 at Solar metallicities, see, e.g.,][]{Davies13,choi+16}.  Despite the robust detections of several RSGs discovered as progenitor stars to SN\,II and with inferred luminosities from $\log(L/L_{\odot})=4.3$--5.2, no sources have been claimed with luminosities exceeding this value.  

The mismatch between the luminosities of the observed population of SN\,II progenitor stars and the overall field population of RSGs led to claims that there are ``missing'' high-mass SN progenitor stars \citep[the ``red supergiant problem'';][]{Smartt15}.  Although this phenomenon could arise at random due to the limited number of SN\,II progenitor stars \citep[][but see also \citealt{Kochanek20} and \citealt{Davies20b}]{Davies20a}, our limited understanding of the late-stage RSG evolution and thus the true spectral shape of these stars \citep{Walmswell12,Davies21,Davies22}, the increasing incidence of close binary interactions among high-mass stars and their evolution into the progenitors of stripped-envelope SNe \citep{Smith11,Zapartas21}, or some combination of the above, there are theoretical expectations and observational evidence that some $\log(L/L_{\odot})=5.2$--5.6 RSGs do not explode as SNe.  These stars tend to have highly compact oxygen cores \citep{Ertl16,Sukhbold16}, which disfavor successful SN explosions and suggest they may instead collapse directly into black holes with at most a weak explosion \citep{OConnor11,Piro13,Fernandez18}.  This phenomenon was invoked to explain a ``disappearing'' 25~$M_{\odot}$ RSG in NGC\,6946 \citep{adams+16} and more recently a 20~$M_{\odot}$ RSG in M31 \citep{De25}.  Systematic searches for these disappearing stars can place upper limits on the fraction of RSGs that produce these ``failed'' SNe \citep{Kochanek14,Farrell20,Byrne22}, while the population of RSGs observed as SN\,II progenitor stars can be better characterized to explain the remainder of the terminal RSG population.

In particular, the impact of circumstellar extinction on direct imaging of SN progenitor stars is difficult to quantify owing to the lack of deep, multi-band infrared (IR) imaging.  A limited number of SN\,II progenitor stars are detected in ground-based IR, {\it HST} Near Infrared Camera and Multi-Object Spectrometer (NICMOS) or Wide Field Camera 3 (WFC3) IR, or {\it Spitzer} Infrared Array Camera (IRAC) images \citep[e.g., SNe\,2005cs, 2012aw, 2017eaw, 2022acko, 2023ixf, and 2024ggi;][]{Li06,Kochanek12,Kilpatrick18:17eaw,VanDyk23a,Kilpatrick23,Xiang24}.  In modeling the SEDs of some of these sources, they appear significantly cooler than expectations for a bare RSG photosphere.  Even accounting for expectations that RSGs evolve toward later spectroscopic types as they approach SN \citep[][]{Davies18}, standard RSG models cannot account for several sources with blackbody temperatures significantly cooler than 3000~K.  Their SEDs imply the presence of a photosphere at radii significantly larger than even the most extreme RSGs \citep[i.e., larger than 2000~$R_{\odot}$ as in][]{Davies07} and thus the existence of a dense, optically-thick shell of circumstellar material (CSM).  The composition of these shells produced at such late evolutionary phases, the mass-loss rates they imply from the progenitor star, and the total circumstellar extinction they impart on the underlying star can be partly constrained from early-time light curves and ``flash spectroscopy'' of SNe \citep{gal-yam+14,khazov+15,Jacobson-Galan24}, but the connection to progenitor stars is only now being studied via exceptional events such as SN\,2023ixf \citep{Kilpatrick23,Jencson23,Niu23,Pledger23,Soraisam23,Neustadt23,VanDyk23}.

Detailed SED modeling of the SN\,2023ixf progenitor star implies a total optical extinction of as much as $A_{V}\approx5$~mag toward the underlying star \citep{Kilpatrick23}, significantly in excess of previous treatment of circumstellar extinction from normal RSG winds where even the highest mass-loss rates were assumed to yield dense CSM with line-of-sight extinction of $A_{V}<1$~mag \citep[as in][]{Walmswell12}.  However, IR detections of nearby counterparts to other SNe, where emission from the star is reprocessed through these optically thick shells of CSM, are needed to better constrain the total amount of light from the star.  With detections at longer mid-IR wavelengths, the composition of this material can also be constrained from broadband silicate and polycyclic aromatic hydrocarbon (PAH) emission similar to those observed in the spectra of Milky Way and Local Group RSGs \citep{verhoelst+09,DeWit23}.

In this paper, we present pre-explosion {\it JWST} and {\it HST} imaging of the nearby SN\,II 2025pht discovered in NGC\,1637 on 29 June 2025 by the All-Sky Automated Survey for Supernovae\footnote{SN\,2025pht is also called ASASSN-25cw.} \citep[ASAS-SN;][]{Shappee14,Kochanek17,2025TNSTR2463....1S}.  Using a post-explosion {\it HST} image of SN\,2025pht, we demonstrate that this event has a single, credible counterpart in pre-explosion imaging and is detected at wavelengths from 0.8--8\,$\mu$m in {\it HST}/WFPC2, {\it JWST}/NIRCam, and {\it JWST}/MIRI from $\approx$24--0.7~yr prior to discovery.  In Section~\ref{sec:progenitor}, we model the photometry using blackbody and realistic stellar SED models.  In Section~\ref{sec:discussion}, we then place the SN\,2025pht progenitor candidate in the context of other SN\,II progenitor stars and the implication for modeling of other SN progenitor stars that lacked such detailed mid-IR detections.

Throughout this paper, we adopt a Milky Way line-of-sight reddening of $E(B-V)=0.036$~mag to SN\,2025pht derived from \citet{Schlafly11}.  We also use the Cepheid distance of 12.03$\pm$0.39~Mpc to NGC\,1637 from \citet{Saha06} and a redshift to NGC\,1637 of $z=0.002392$ from \citet{Springbob05}.

\section{Observations of SN~2025\lowercase{pht} and Its Progenitor Candidate}\label{sec:observations}

\subsection{Post-Explosion {\it Hubble Space Telescope} Imaging}\label{sec:post_hst}

We observed the site of SN\,2025pht on 31 Jul 2025 with the Wide Field Camera 3 in the F336W band.  We used 16$\times$70~s (1120~s total) exposures over a single orbit, reading out only the UVIS2-C1K1C subarray and using a 4-point box dither pattern around our target to maximize open shutter time.  Downloading all of the {\tt flc} frames, we processed these data using {\tt hst123}, including frame-to-frame alignment with {\tt TweakReg} \citep{tweakreg}, drizzling using {\tt astrodrizzle} \citep{drizzlepac}, and photometry of sources detected in the image frame with {\tt dolphot} \citep{dolphot}.  SN\,2025pht is clearly detected in the output {\tt dolphot} catalog, with a brightness of $m_{\rm F336W}=19.06 \pm 0.01$~mag\footnote{All photometry reported throughout this paper is given on the AB magnitude system \citep{Oke83}.}.

\subsection{Spectroscopy of SN~2025\lowercase{pht}}\label{sec:spec_obs}

SN\,2025pht was observed with the Low-Resolution Imaging Spectrograph \citep[LRIS;][]{LRIS} on the 10~m Keck~I telescope on 27.64 July 2025. The 600/4000 grism was used on the blue channel, along with the 1200/7500 grating on the red channel. This setup achieves a nominal dispersion of 0.63~\AA\ mm$^{-1}$ between 3140 and 5640~\AA, and 0.4~\AA\ mm$^{-1}$ between 5680 and 7340~\AA.

To reduce the LRIS data, we used the {\tt UCSC Spectral Pipeline}\footnote{\url{https://github.com/msiebert1/UCSC\_spectral\_pipeline}} \citep{Siebert20}, a custom data-reduction pipeline based on procedures outlined by \citet{Foley03}, \citet{Silverman2012}, and references therein.  The two-dimensional (2D) spectra were bias-corrected, flat-field corrected, adjusted for varying gains across different chips and amplifiers, and then trimmed to the data section.  One-dimensional spectra were extracted using the optimal algorithm \citep[see][]{Horne86}.  The spectra were wavelength-calibrated using internal comparison-lamp spectra with linear shifts applied via cross-correlation of the observed night-sky lines in each spectrum to a master night-sky spectrum.  Flux calibration and telluric correction were also performed using standard stars observed at a similar airmass to that of the science exposures. A more detailed description of this reduction process is provided elsewhere \citep{Foley03, Silverman2012, Siebert20}.

\subsection{Pre-Explosion {\it Hubble Space Telescope} Imaging}\label{sec:pre_hst}

The site of SN\,2025pht was observed over several visits by \hst\ from 1994 to 2024 with the Wide Field Planetary Camera 2 (WFPC2) and WFC3/UVIS.  We downloaded the calibrated exposures\footnote{i.e., {\tt c0m} files for WFPC2 and {\tt flc} files for WFC3.} and produced drizzled stacks for each unique visit and filter using {\tt hst123} \citep{hst123} following procedures described in \citet{Kilpatrick23}.  In addition, we produced deep stacks for all WFPC2 F555W and F814W images.

We obtained photometry of point sources in all \hst\ images using {\tt dolphot} \citep{dolphot}.  Using our deep WFPC2/F814W stack as a reference image, we used standard {\tt dolphot} parameters\footnote{Including {\tt FitSky=2} and a final signal-to-noise threshold of {\tt SigFinal=3.5}.} to measure the brightness of faint sources near the site of SN\,2025pht.  We produced a final {\it HST} catalog of point sources within 10\arcsec\ of the reported coordinates of SN\,2025pht with a final signal-to-noise of $\geq$3.5, {\tt dolphot} crowding parameter $<$2, sharpness$^{2}<0.3$, and {\tt dolphot} quality flag $\leq$3.

\subsection{{\it James Webb Space Telescope} Imaging}\label{sec:jwst}

The site of SN\,2025pht was observed by the {\it James Webb Space Telescope} ({\it JWST}) using the Near Infrared Camera (NIRCam) and Mid-Infrared Instrument (MIRI) over two epochs on 5 Feb 2024 and 8 Oct 2024.  We obtained all level-2 images and processed them using our custom pipeline described in \citet{Blanchard25}, including frame-to-frame alignment with {\tt JHAT} \citep{jhat}, mosaicing each band into level-3 images using the {\it JWST} pipeline \citep{Bushouse23}, and photometry of all sources in the level-2 images using {\tt dolphot}.  We applied the same {\tt dolphot} quality cuts to the {\it JWST} catalogs as in our {\it HST} reductions.

\section{Spectral Classification of SN~2025\lowercase{pht}}

\subsection{Spectral Classification and Approximate Phase}\label{sec:spec}

\begin{figure}
    \includegraphics[width=0.49\textwidth]{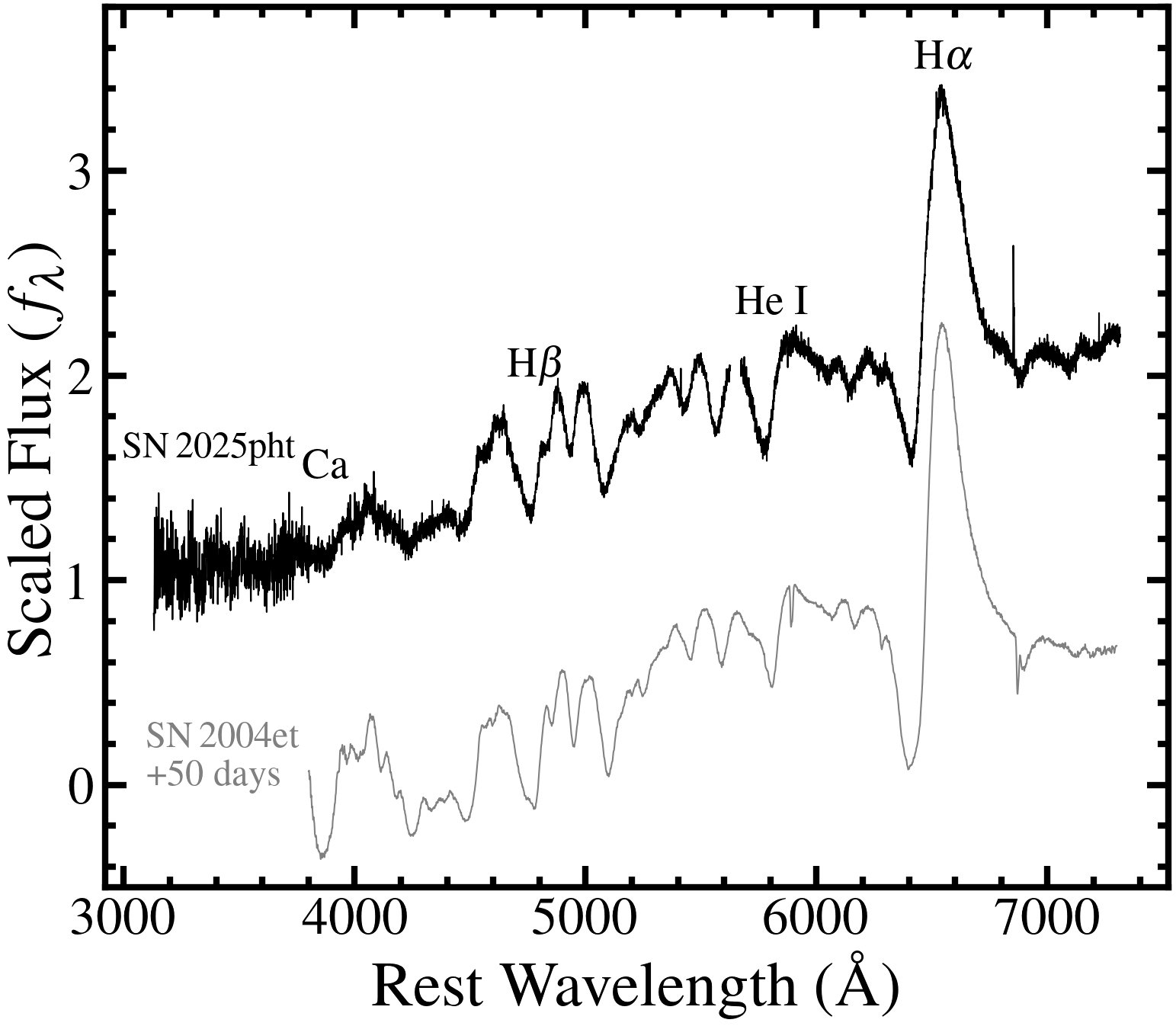}
    \includegraphics[width=0.49\textwidth]{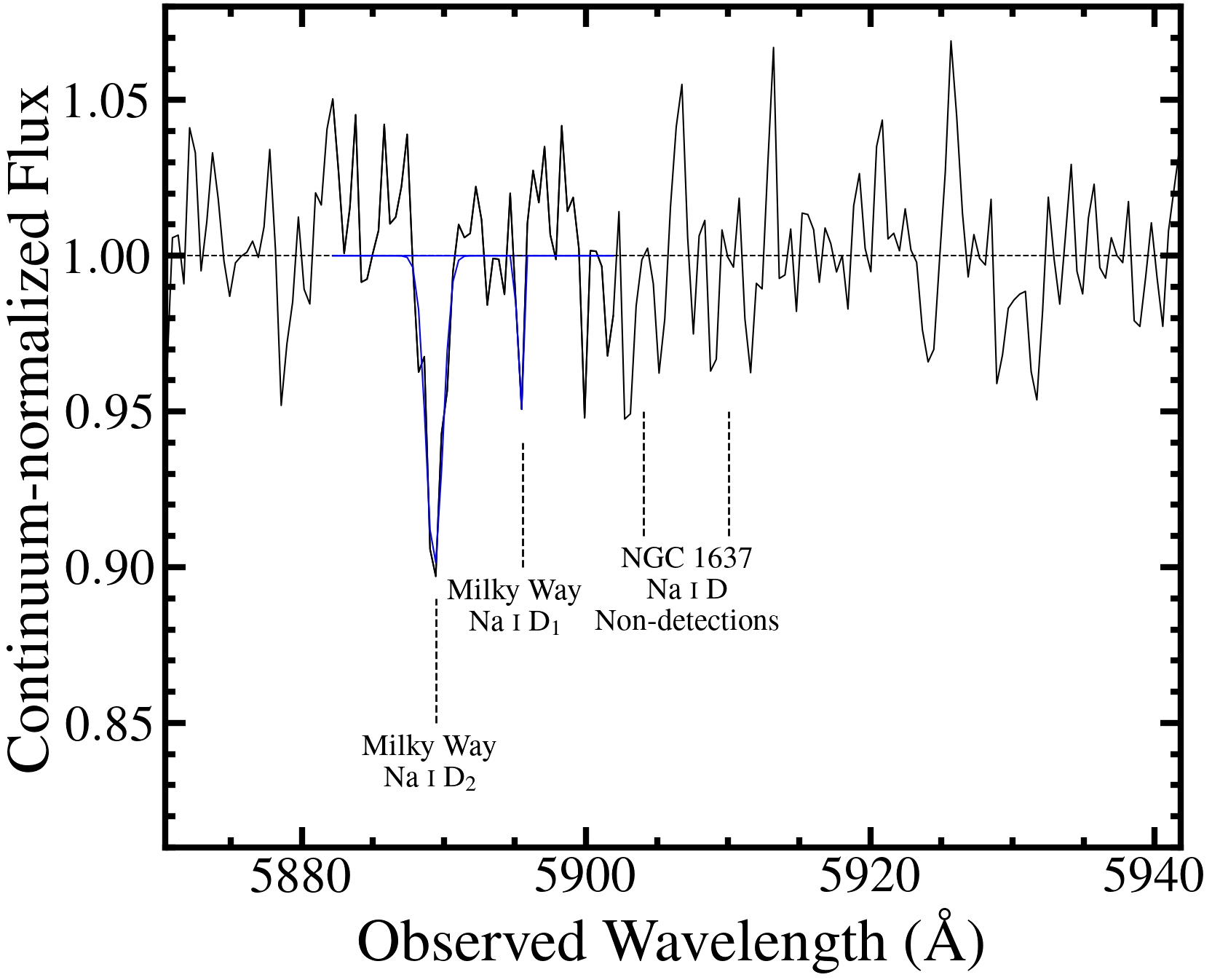}
    \caption{{\it Top}: Our Keck/LRIS spectrum of SN\,2025pht (black) obtained on 27 July 2025 as described in Section~\ref{sec:spec_obs} and compared with a spectrum of SN\,2004et (grey) that is most consistent with our observation (Section~\ref{sec:spec}).  We identify key features that aid in our classification of SN\,2025pht as a typical SN\,II, likely in its plateau phase during this epoch and with a H$\alpha$ velocity derived from the trough of its blueshifted absorption of 6800$\pm$100~km~s$^{-1}$.  {\it Bottom}: A continuum-normalized section of the SN\,2025pht spectrum centered on the rest wavelength of Na\I~D at the redshift of NGC\,1637 ($\approx$5906~\AA).  While we detect the Milky Way components of Na\I~D, we do not detect any Na\I~D absorption from NGC\,1637 leading us to conclude there is an insignificant magnitude of line-of-sight extinction to SN\,2025pht in this galaxy.}\label{fig:spectrum}
\end{figure}

We compare our Keck/LRIS spectrum of SN\,2025pht to those of other SN\,II using {\tt SNID} \citep{snid}, finding the best matches to those of SN\,2004et during the plateau phase.  In particular, the best-fitting SN\,2004et spectrum was from 11 Nov 2004 as originally presented in \citet{Sahu06}, which corresponds to a phase relative to their derived explosion date of $\approx50$~days.  Overall, the broad features of hydrogen and helium are well matched as shown in Figure~\ref{fig:spectrum}.

In order to validate the approximate phase more quantitatively, we measure a H$\alpha$ velocity from the trough of the blueshifted P-Cygni absorption feature of 6800$\pm$100~km~s$^{-1}$ for SN\,2025pht.  This is in good agreement with the measurement from SN\,2004et in the best-fitting spectrum of 7000$\pm$300~km~s$^{-1}$ from \citet{Sahu06}.  We conclude that SN\,2025pht is a SN\,2004et-like Type\,II-P SN that was likely in its plateau phase during our 27 July 2025 observation.

\subsection{Constraints on Host Galaxy Reddening Derived from Na\textsc{i}~D}\label{sec:naid}

We use our high-resolution Keck spectrum of SN\,2025pht to derive a line-of-sight extinction based on Na\I~D lines.  The equivalent width of the Na\I~D doublet is correlated with extinction, and stronger Na\I~D absorption features are observed when extinction and reddening are more pronounced, larger extinction values.  Relations that correlate Na\I~D absorption with extinction and reddening have been derived using spectra and light curves of quasars and SNe \citep[see, e.g.,][]{Poznanski12,Phillips13,Galbany16,Stritzinger18}.

We show the continuum-normalized spectrum centered around the rest wavelength of Na\I~D in NGC\,1637 in Figure~\ref{fig:spectrum}.  The Milky Way Na\I~D component is clearly observed at the appropriate rest wavelengths of 5889.95~\AA\ and 5895.92~\AA\ for Na\I~D$_{2}$ and Na\I~D$_{1}$, respectively, and we derive equivalent widths for both components of EW$_{{\rm Na}\I~{\rm D}_{2}}=0.15\pm0.04$~\AA\ and EW$_{{\rm Na}\I~{\rm D}_{1}}=0.05\pm0.02$~\AA.  However, we do not observe any significant Na\I~D emission at the redshift of NGC\,1637.  If we force a fit at the wavelength of Na\I~D$_{2}$ at this redshift (i.e., with the centroid fixed at 5904.04~\AA) and the line width fixed to that of the Milky Way component ($\approx$0.59~\AA), the best-fitting equivalent width is 0.00$\pm$0.02~\AA, from which we infer that these lines are not detected.  From the relations of \citet{Poznanski12}, this implies a line-of-sight host reddening of $E(B-V)<0.02$~mag at the 3$\sigma$ level.  Thus we do not apply any host extinction to the pre-explosion counterpart of SN\,2025pht in the modeling below.

\section{The Progenitor Candidate of SN~2025\lowercase{pht}}\label{sec:progenitor}

\subsection{Alignment between Pre- and Post-Explosion Imaging and Photometry of the SN~2025pht Progenitor Candidate}\label{sec:alignment}

We aligned our {\it HST} image of SN\,2025pht to the pre-explosion {\it JWST}/NIRCam F150W image using {\tt JHAT} \citep{jhat}.  Using 36 sources common to both image frames, we applied a shift between the two image frames that minimized the root-mean square dispersion between these two image frames.  {\tt JHAT} reported a final dispersion of 0.030\arcsec, which we adopt as the total uncertainty in the astrometric alignment between the pre- and post-explosion imaging.

Following this analysis, we compared the centroids of all sources detected by {\tt dolphot} in the {\it HST} and {\it JWST} images around the locations shown in Figure~\ref{fig:images}.  SN\,2025pht is clearly visible in the F336W imaging based on comparison to the pre-explosion F336W image (Figure~\ref{fig:images}) with negligible astrometric uncertainty on its centroid located at $\alpha_{\rm J2000} = 4^{\rm h}41^{\rm m}28^{\rm s}.8767$ and $\delta_{\rm J2000} = -2^{\circ}51^{\prime}55^{\prime\prime}.881$ in the {\it Gaia}-aligned {\it HST} image frame.  Comparing to sources in our {\it JWST} catalog, there is a single, bright, point-like object in F150W whose centroid is located at $\alpha_{\rm J2000} = 4^{\rm h}41^{\rm m}28^{\rm s}.8758$ and $\delta_{\rm J2000} = -2^{\circ}51^{\prime}55^{\prime\prime}.871$, implying an offset of 0.017\arcsec\ (0.6$\sigma$).  We conclude that these sources are astrometrically coincident to the level of precision of our frame-to-frame alignment and the F150W source is consistent with being the pre-explosion counterpart to SN\,2025pht.

\begin{figure*}
    \includegraphics[width=\textwidth]{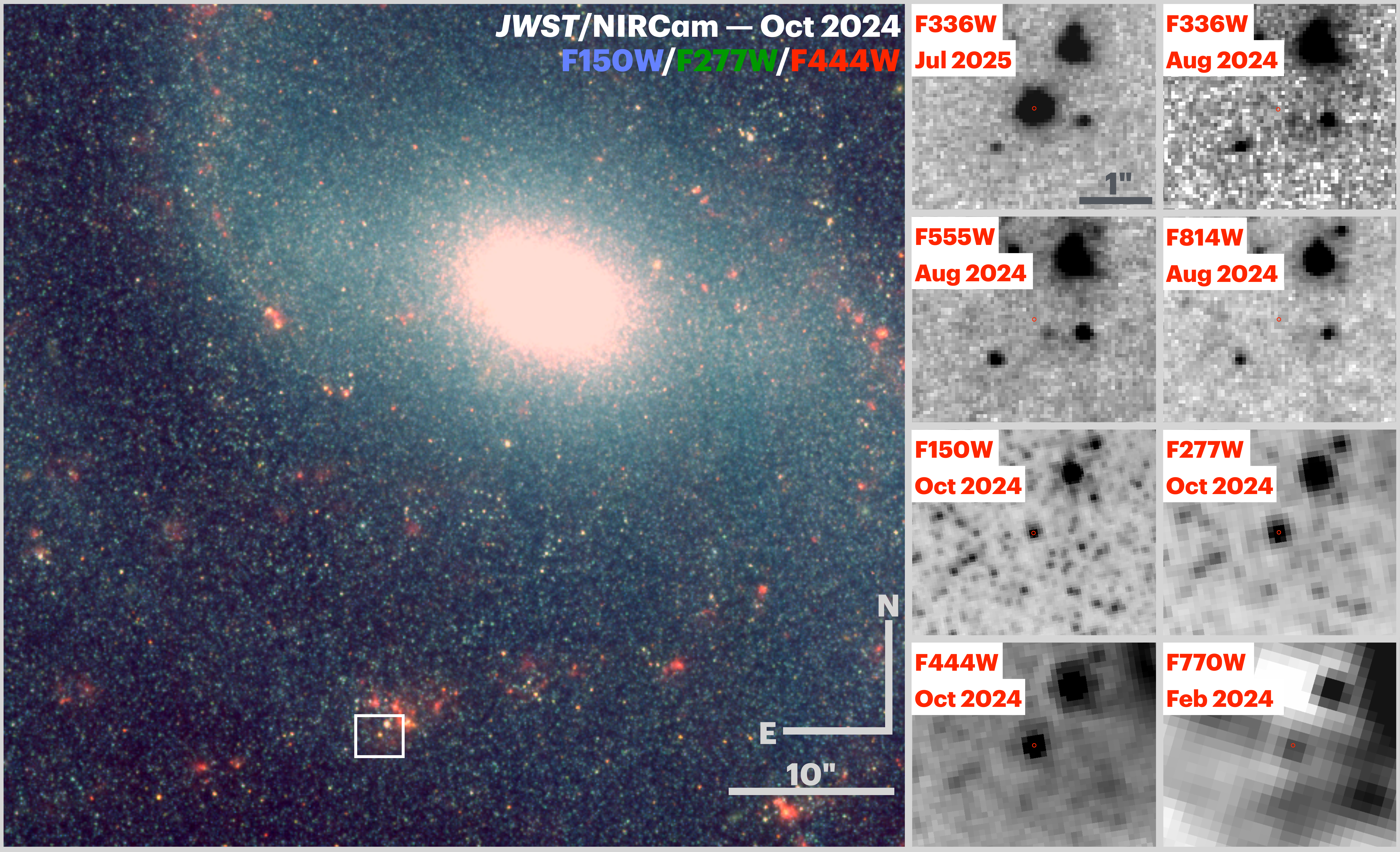}
    \caption{{\it Left}: A color image of NGC\,1637 constructed from F150W (blue), F277W (green), and F444W (red) images obtained with {\it JWST}/NIRCam on 8 Oct 2024 showing a 1.0$\times$0.9~arcmin$^{2}$ portion of the galaxy.  We highlight a 2.7$\times$2.4~arcsec$^{2}$ region centered on the site of SN\,2025pht with a white box toward the bottom of the frame.  {\it Right}: a series of grayscale frames showing {\it HST}/WFC3 F336W imaging of SN\,2025pht obtained on 31 July 2025 with the centroid of the SN marked in red.  We show the same region obtained by {\it HST}/WFC3 in F336W, F555W, and F814W obtained on 3 Aug 2024, {\it JWST}/NIRCam in F150W, F277W, and F444W on 8 October 2024, and by {\it JWST}/MIRI on 5 Feb 2024.  We show the location of the SN with the same red circle demonstrating that there is a credible counterpat detected in all of the {\it JWST} images as discussed in Section~\ref{sec:alignment}.}\label{fig:images}
\end{figure*}

This source is detected across every pre-explosion {\it JWST}/NIRCam image, all pre-explosion WFPC2/F814W imaging from 2001 (Figure~\ref{fig:f814w}), and in {\it JWST}/MIRI F770W as shown in Figure~\ref{fig:images}.  In every other image, we set an approximate limiting magnitude using a forced aperture at the site of SN\,2025pht set to the 95\% encircled energy radius in each band with the appropriate zero point and aperture correction.  All detections and upper limits across each image is reported in Table~\ref{tab:hst} and Table~\ref{tab:jwst} for {\it HST} and {\it JWST}, respectively.

\begin{figure}
    \includegraphics[width=0.49\textwidth]{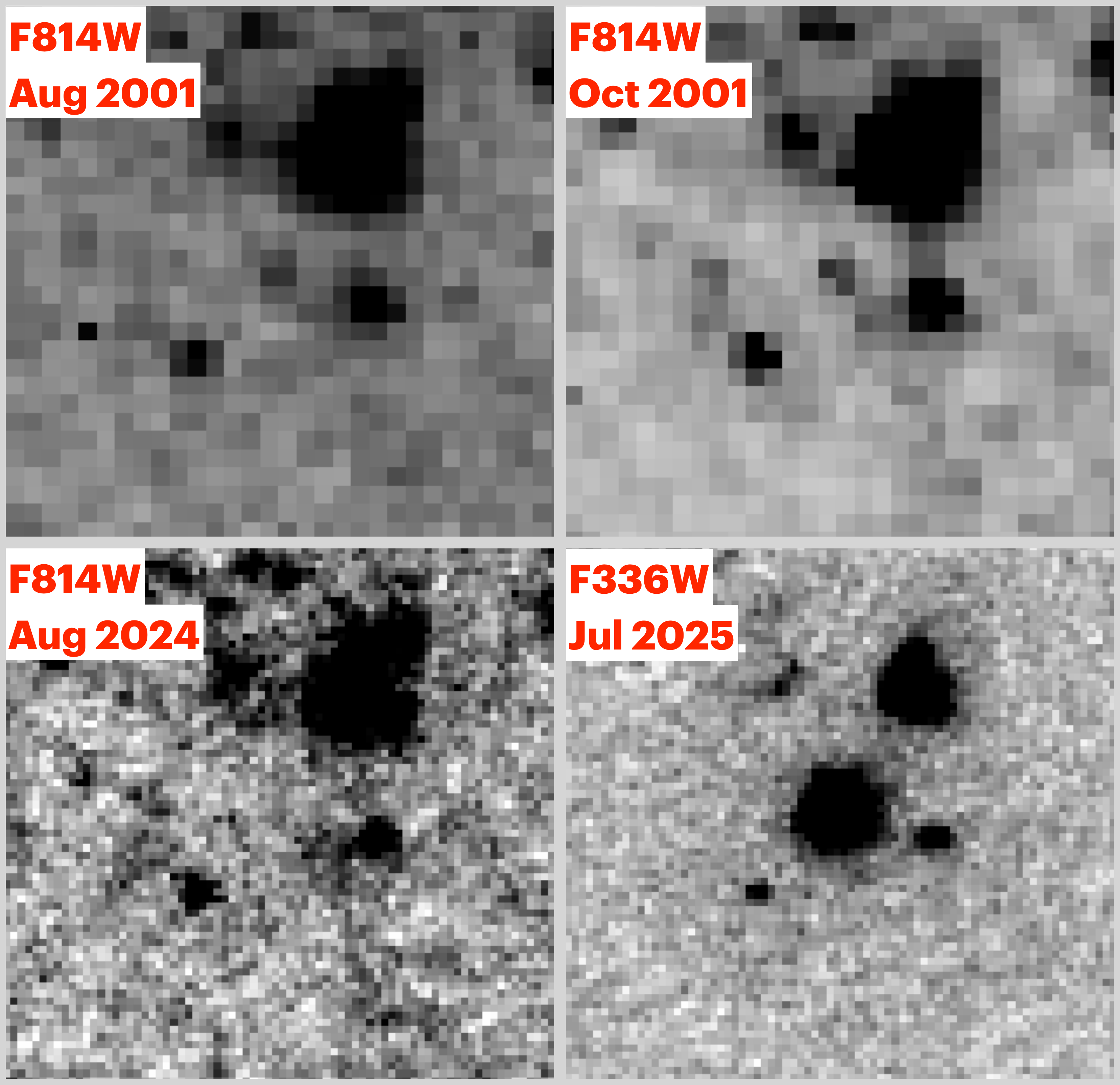}
    \caption{A series of {\it HST}/WFPC2 ({\it upper}) and WFC3 ({\it lower}) images in F814W and F336W showing the detection of SN\,2025pht ({\it lower right}) and the detection of its counterpart from 2001 and eventual disappearance in 2024, demonstrating that the pre-explosion counterpart had some long time-baseline variability over this 23~yr span.  We discuss the significance of this measurement in Section~\ref{sec:variability}.}\label{fig:f814w}
\end{figure}

We also consider the possibility that the pre-explosion counterpart to SN\,2025pht was detected due to chance coincidence and set an approximate probability of chance coincidence using the following method.  There are 417 sources within a 3\arcsec\ radius of the site of SN\,2025pht and detected at $>$5$\sigma$ in the F150W image that also pass the quality cuts we describe to the {\tt dolphot} catalog in Section~\ref{sec:jwst}.  Thus for any source found near the location of SN\,2025pht, there is roughly a 4\% chance that it would be detected within 1$\sigma$ of the location of an arbitrary source.  We consider this to be a small but non-negligible chance that the counterpart was detected by chance, although we do not impose any other cuts (e.g., on color, brightness, or counterparts in other catalogs) that could be used to simulate the detection of a likely SN\,II counterpart.  Furthermore, we can validate whether the counterpart is the SN progenitor star through its eventual disappearance \citep[similar to][]{VanDyk23c}.

Throughout the remainder of this paper, we proceed under the assumption that the detected counterpart is the progenitor star of SN\,2025pht and its photometry is accurately described by the values in Tables~\ref{tab:hst} and \ref{tab:jwst}.  However, we note that the F164N, F212N, and F405N detections ([Fe\II], H$_{2}$, and Brackett-$\alpha$ emission), while also coincident with the pre-explosion counterpart, are moderately brighter than the wide-band detections that are closest in wavelength (F150W, F200W, and F444W, respectively).  The counterpart is also detected in F187N (Paschen-$\alpha$), but this detection is nominally fainter than in F200W, implying a low line flux.  Assuming a relatively flat continuum at these wavelengths with the rest of the emission dominated by spectral lines, we derive integrated line fluxes of $f_{[{\rm Fe}\II]}=2.4$~$\mu$Jy, $f_{{\rm H}_{2}}=3.1$~$\mu$Jy, and $f_{{\rm Br}\alpha}=0.61$~$\mu$Jy.  This emission could plausibly arise from the circumstellar environment around the progenitor star \citep[similar to the molecular outflows observed in RSGs such as NML Cyg;][]{Beck25} or an unassociated nebula.  However, as the detailed radiative transfer calculations that can accurately model this line flux are beyond the scope of this paper, we do not use any narrow-band photometry in the analysis below.

\begin{deluxetable}{ccccc}
\tablecaption{{\it HST} Photometry of the SN\,2025pht Progenitor Candidate}
\tablehead{
\colhead{MJD} &
\colhead{Instrument} &
\colhead{Band} &
\colhead{Magnitude} &
\colhead{Uncertainty} \\
&
&
&
\colhead{(AB mag)} &
\colhead{(mag)}
}
\startdata
49606.190 & WFPC2 & F606W    & $>$24.87 & -- \\
52133.332 & WFPC2 & F450W    & $>$25.09 & -- \\
52133.342 & WFPC2 & F814W    & 25.23 & 0.34 \\
52154.191 & WFPC2 & F814W & 25.16 & 0.24 \\
52162.951 & WFPC2 & F814W & 25.06 & 0.23 \\
52178.531 & WFPC2 & F814W & 25.09 & 0.24 \\
52181.874 & WFPC2 & F814W & 25.15 & 0.24 \\
52184.671\tablenotemark{a} & WFPC2 & F555W    & $>$26.61 & -- \\
52200.658 & WFPC2 & F814W & 25.20 & 0.24 \\
52213.625 & WFPC2 & F814W & 25.06 & 0.23 \\
60520.061 & WFC3/UVIS & F657N    & $>$24.55 & -- \\
60525.579 & WFC3/UVIS & F814W    & $>$26.05 & -- \\
60525.584 & WFC3/UVIS & F438W    & $>$26.58 & -- \\
60525.590 & WFC3/UVIS & F336W    & $>$25.31 & -- \\
60525.596 & WFC3/UVIS & F275W    & $>$26.85 & -- \\
60525.605 & WFC3/UVIS & F555W    & $>$26.59 & -- \\
\enddata
\tablecomments{The MJD corresponds to the time at the start of each {\it JWST} image.  All photometry is on the AB magnitude scale, and limiting magnitudes are provided at 3$\sigma$ as described in Section~\ref{sec:alignment}.}\label{tab:hst}
\tablenotetext{a}{This WFPC2 epoch is a deep stack of 12$\times$2200~s F555W exposures obtained from 2 Sep 2001 to 31 Oct 2001.  The MJD provided is the average of the start dates for each exposure.}
\end{deluxetable}

\begin{deluxetable}{ccccc}
\tablecaption{{\it JWST} Photometry of the SN\,2025pht Progenitor Candidate}
\tablehead{
\colhead{MJD} &
\colhead{Instrument} &
\colhead{Band} &
\colhead{Magnitude} &
\colhead{Uncertainty} \\
&
&
&
\colhead{(AB mag)} &
\colhead{(mag)}
}
\startdata
60345.633 & MIRI & F0770W & 21.25 & 0.11 \\     
60345.639 & MIRI & F2100W & $>$17.04 & -- \\
60345.679 & NIRCam & F150W & 22.12 & 0.01 \\
60345.679 & NIRCam & F300M & 21.13 & 0.01 \\
60345.688 & NIRCam & F187N & 21.94 & 0.02 \\
60345.688 & NIRCam & F335M & 20.96 & 0.01 \\
60591.272 & NIRCam & F187N & 21.99 & 0.02 \\
60591.272 & NIRCam & F405N & 20.84 & 0.01 \\
60591.299 & NIRCam & F164N & 21.70 & 0.02 \\
60591.299 & NIRCam & F335M & 21.01 & 0.01 \\
60591.326 & NIRCam & F212N & 21.26 & 0.01 \\
60591.326 & NIRCam & F360M & 20.89 & 0.01 \\
60591.352 & NIRCam & F150W & 22.31 & 0.01 \\
60591.352 & NIRCam & F444W & 20.88 & 0.01 \\	
60591.370 & NIRCam & F200W & 21.60 & 0.01 \\
60591.370 & NIRCam & F277W & 21.43 & 0.01 \\
\enddata
\tablecomments{The MJD corresponds to the time at the start of each {\it JWST} image.  All photometry is on the AB magnitude scale, and limiting magnitudes are provided at 3$\sigma$ as described in Section~\ref{sec:alignment}.}\label{tab:jwst}
\end{deluxetable}

\subsection{Evidence for Variability from the SN~2025\lowercase{pht} Progenitor Candidate}\label{sec:variability}

Motivated by previous studies of extreme variability in SN\,II progenitor candidates \citep[e.g., SN\,2020tlf and 2023ixf, which exhibited some level of pre-explosion optical or mid-IR variability;][]{Jacobson-Galan22,Kilpatrick23,Jencson23,Soraisam23}, we quantify whether the SN\,2025pht counterpart exhibits similar evidence for detectable variability in the optical and mid-IR.  We have three avenues of investigation to quantify this for the SN\,2025pht counterpart: i) the seven epochs of WFPC2/F814W photometry from 12 Aug to 31 Oct 2001, ii) the difference between these epochs and the 3 Aug 2024 WFC3/F814W photometry, and iii) the two epochs of NIRCam/F150W and F335M photometry on 5 Feb and 8 Oct 2024.

For the former, the low signal-to-noise ratio of the individual WFPC2/F814W detections (e.g., in Figure~\ref{fig:images}) preclude any strong statement on the intrinsic variability of the underlying source.  Over this 80~day period, the source was less variable than the peak-to-peak difference combined with the individual statistical uncertainties on each measurement, or $\lesssim$0.3~mag.  On the other hand, the difference between the detection in 2001 and non-detection in 2024 is more extreme, with a 0.84~mag (factor of 2.2) decline in brightness.  While there are examples of stars with similar levels of variability at these wavelengths \citep[$\alpha$~Ori had similar peak-to-peak variability from 2017--2019 at these wavelengths;][]{Taniguchi22}, the 0.84~mag decline is only a lower limit on our knowledge of the counterpart's variability and already makes the counterpart among the most extreme $M_{I}<-5$~mag variables observed on timescales of decades \citep[e.g., those in M51 in][]{Conroy18}.

Finally, the level of variability measured in F150W (+0.19~mag) and F335M (+0.05~mag) is weak compared with extreme IR sources such as the SN\,2023ixf counterpart.  Although we only measure this source in two epochs and cannot firmly conclude that we have captured the full range of changes in this source, we do not see any strong evidence for changes on this timescale of 246~days \citep[roughly 1/4 the period derived from variability in SN\,2023ixf;][]{Kilpatrick23}.

For the purposes of our analysis and given the lack of strong variability in the observations from 2024, we consider the $m_{\rm F814W}>26.05$~mag WFC3/UVIS limit to be near-contemporaneous with the NIRCam and MIRI detections.  Therefore, in the modeling of this sources SED below, we adopt this limit in F814W as opposed to the prior detections from 2001.

\subsection{The Spectral-Energy Distribution of the SN~2025pht Progenitor System}\label{sec:sed}

\subsubsection{Blackbody Spectral Energy Distributions}

We first fit broadband photometry of the SN\,2025pht pre-explosion counterpart with a blackbody model.  We forward model SEDs of a fixed luminosity ($\log(L/L_{\odot})$) and effective temperature ($T_{\rm eff}$) using realistic bandpasses for each {\it JWST} and {\it HST} filter and using {\tt synphot} \citep{synphot} to derive the expected brightness of the counterpart.  We adopt the distance and redshift given above and assume a \citet{Fitzpatrick99} extinction law for Milky Way dust via the {\tt extinction} module \citep{extinction}.  As described in Section~\ref{sec:naid}, we assume zero host extinction.

We fit this model using {\tt emcee} \citep{emcee} by optimizing the likelihood function:

\begin{small}
\begin{equation}
\begin{split}
    &\log \mathscr{L}(\theta) = \\
    &-\frac{1}{2} \sum_{i}^{}\left(\frac{(m_{i, {\rm obs}} - m_{i, {\rm model}}(\theta))^{2}}{\sigma_{i}^{2} + \sigma_{\rm sys}^{2}} + \log(2 \pi (\sigma_{i}^{2} + \sigma_{\rm sys}^{2})) \right) \label{eqn:likelihood}
\end{split}
\end{equation}
\end{small}

\noindent where $m_{i, {\rm obs}}$ is an observed magnitude and $\sigma_{i}$ is its corresponding uncertainty, $m_{i, {\rm model}}$ is a magnitude derived from our forward model for input model parameters $\theta$ in that observation's bandpass, and $\sigma_{\rm sys}$ is a systematic uncertainty term that we fit for to account for additional variance in the data (e.g., due to unmodeled variability) or other systematic uncertainties in the model.

We run our blackbody model for 5,000 steps after an initial burn-in of 1000 steps and using 100 walkers, and from the samples on our {\tt emcee} chains, we report the median, 16th and 84th percentile model parameters from the output posteriors as our best-fitting and 1-$\sigma$ uncertainty range on each parameter.  For this run, the best-fitting parameters are $\log(L/L_{\odot})=4.96\substack{+0.10\\-0.09}$, $T_{\rm eff}=1470\pm40$~K, and $\sigma_{\rm sys}=0.212\substack{+0.060\\-0.046}$~mag.  The luminosity contains the added systematic uncertainty on the distance, although we do not include these values in our fit since the uncertainty is correlated across the SED and our model magnitudes are simply scaled to the correct distance.

Similar to SN\,2023ixf \citep{Kilpatrick23,Jencson23}, SN\,2022acko \citep{VanDyk23a}, and SN\,2024ggi \citep{Xiang24}, all of which had a pre-explosion counterpart detected in multiple bands including at $>$1~micron, the blackbody temperature implied by the pre-explosion counterpart photometry is much cooler than any known supergiant photosphere.  We therefore infer that there must be an extended mass of CSM beyond the SN\,2025pht progenitor star in which a cooler photosphere forms, roughly at the derived blackbody radius of $\approx$3700~$R_{\odot}$.  Assuming the counterpart is a RSG with a total $\log(L/L_{\odot})=4.96$ \citep[$M_{\rm ZAMS}\approx$14.5~$M_{\odot}$ assuming a MIST single-star model;][]{choi+16}, the photosphere would be 5.3 times as large as the underlying star, suggesting an optically-thick, dusty source that forms in a dense RSG wind.  We therefore turn to a more sophisticated SED to constrain the properties of both this material and the underlying star.

\begin{figure*}
    \includegraphics[width=\textwidth]{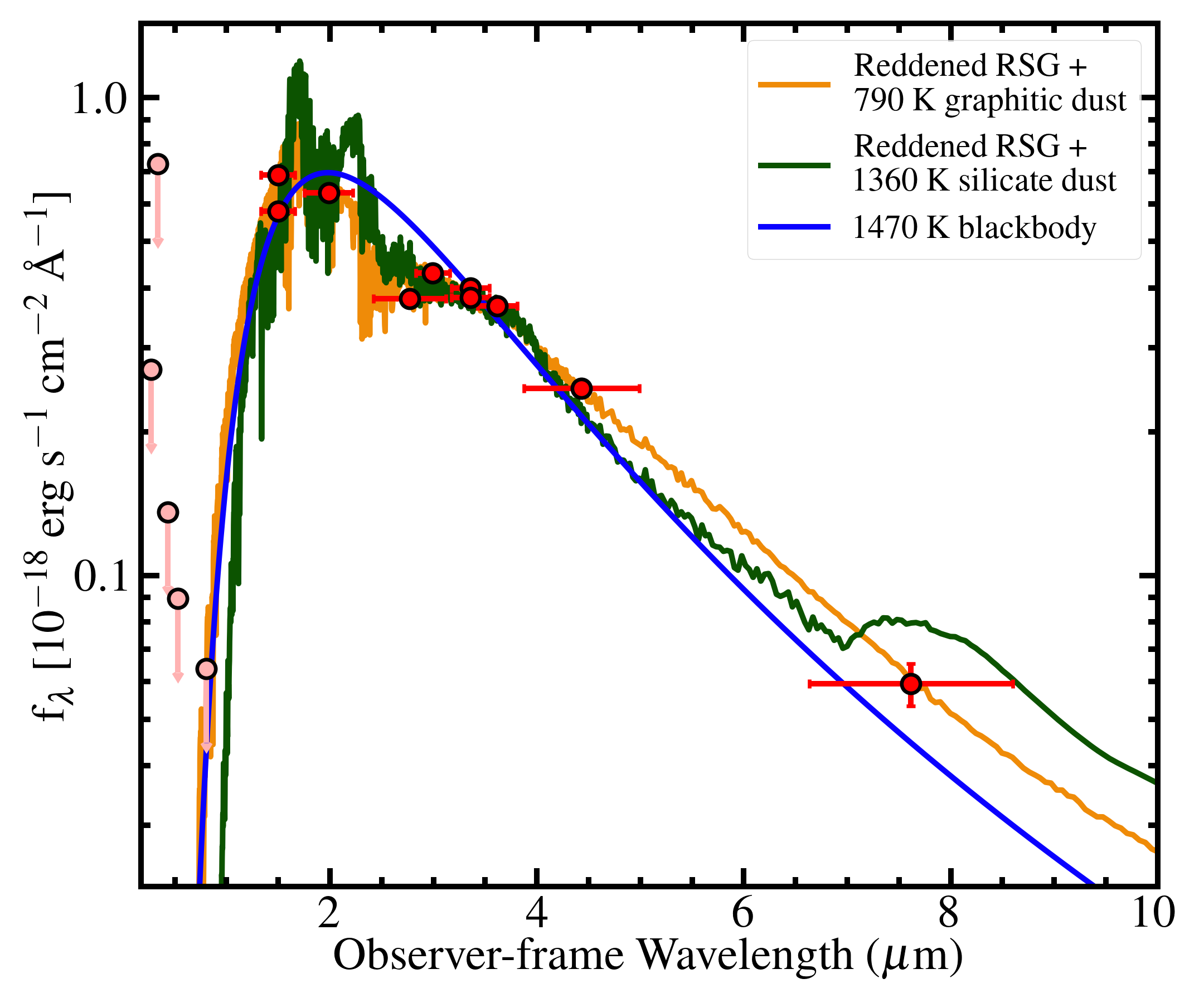}
    \caption{Our best-fitting SEDs to the 2024 detections of the SN\,2025pht counterpart as discussed in Section~\ref{sec:sed}.  All SEDs correspond to a source with $\log(L/L_{\odot})\approx5.0$.  Motivated by the cool blackbody fit, we attempt fits to MARCS SEDs of RSGs reddened via dust prescriptions using {\tt dusty} as described in Section~\ref{sec:rsg_sed}.  While a purely graphitic dust prescription captures the overall shape of the SED out to 8~$\mu$m, the silicate model overpredicts this measurement, leading us to conclude that the SN\,2025pht counterpart is consistent with a graphitic dust model.}\label{fig:sed}
\end{figure*}

\subsubsection{Red Supergiant Spectral-Energy Distributions}\label{sec:rsg_sed}

We consider SED models based on realistic spectra of RSGs derived from MARCS \citep{gustaffson+08} and fully passed through the radiative transfer code {\tt dusty} \citep{dusty} to describe the SN\,2025pht progenitor star.  We take MARCS SEDs corresponding to RSGs with effective temperatures $T_{\rm eff}=2600$--5000~K and scaled to luminosities from $\log(L/L_{\odot})=3.0$--5.8.  All of these SEDs were calculated at Solar metallicity, which is the closest available model to the global metallicity in NGC\,1637 of $\log [{\rm Fe/H}] =-0.11$ \citep{Galbany16b}.  We then model a dust-obscured RSG by processing each SED through a model dust shell containing dust with a grain size ($a$) distribution of $dn/da \propto a^{-3.5}$ \citep[i.e., a MRN distribution;][]{Mathis1977} from $a=0.005$--0.25~$\mu$m.  The shell has a fixed effective temperature from 500--1500~K and an optical depth at 5500~\AA\ of $\tau_{V}=0.01$--15.0.  The geometry of the dust shell follows a density profile with $\rho(r) \propto r^{-2}$ and an outer radius ($R_{\rm out}$) fixed to $2 R_{\rm in}$, where the inner radius $R_{\rm in}$ is degenerate with $\tau_{V}$ and is not explicitly defined.  We note that, based on our estimate of the phase of SN\,2025pht above and the fact that there was no appreciable Na\I~D absorption at +50~days, we can set an approximate upper limit on the size of the dust shell.  We assume that the initial ejecta velocity was $\approx$10,000~km~s$^{-1}$ \citep[which is high for typical SN\,II;][]{Martinez22}, implying that at a radius of at most $\approx$62,000~$R_{\odot}$ there was no circumstellar extinction and thus we can limit possible values of $R_{\rm out}$.

Finally, we consider two compositions that define the dust grain abundances and their corresponding opacities from \citet{Draine1984}: a pure graphite and a pure silicate model.  We fit these dust composition models separately, but otherwise we allow the model parameters ($\log(L/L_{\odot})$, $T_{\rm eff}$, $T_{\rm dust}$, $\tau_{V}$) to vary within the bounds described above.

Following a similar fitting method to that of the blackbody model, we ran our graphitic model for 25,000 steps with a 5000 step burn-in with 100 walkers and found a best-fitting model with $\log(L/L_{\odot})=5.00\substack{+0.09\\-0.08}$, $T_{\rm eff}=3030\substack{+210\\-270}$~K and enshrouded with a dust shell with $T_{\rm dust}=790\substack{+40\\-30}$~K and $\tau_{V}=6.7\pm0.4$.  The final systematic variance was $\sigma_{\rm sys}=0.08\substack{+0.03\\-0.02}$~mag.

Similarly, we fit the silicate model to the pre-explosion photometry, but the model did not converge and all of the walkers ended at the edge of the opacity boundary with $\tau_{V}=15$.  We increased this parameter to allow for models with $\tau_{V}=0.01$--100 and performed the same fit again with the best-fitting model exhibiting $\log(L/L_{\odot})=5.01\pm0.09$, $T_{\rm eff}=2800\substack{+450\\-150}$~K, $T_{\rm dust}=1360\substack{+90\\-100}$~K, and $\tau_{V}=22.8\substack{+2.1\\-2.4}$ with $\sigma_{\rm sys}=0.15\substack{+0.05\\-0.04}$~mag (Figure~\ref{fig:sed}).  However, we note that the model predicts a significantly higher MIRI/F770W flux than the observation, which is inconsistent with the detection at the $>$3$\sigma$ level.  This is inconsistency does not occur with the graphitic model, which has a smoother continuum and is well-matched to the MIRI/F770W observation.  We emphasize that there is a clear feature at these wavelengths associated with silicate-rich dust \citep{Draine1984}, which is often used to distinguish between dust compositions in the spectra of RSGs \citep[e.g.,][]{verhoelst+09}.  

Although, superficially, this mismatch between the model and photometry would appear to imply that $\tau_{V}$ is overestimated, when we perform the same silicate-rich model fit using only the NIRCam photometry we derive $\tau_{V}=19.2\substack{+3.1\\-2.7}$ with a similarly prominent silicate feature as shown in Figure~\ref{fig:models}.  This model largely captures the shape of the short-wavelength SED, suggesting that it is primarily the NIRCam photometry driving the high-opacity prediction for silicate-rich dust models.  Similarly, we derive $\tau_{V}=6.3$ for the graphitic model in this NIRCam-only fit, consistent with the value derived above, indicating that this model can accurately predict the MIRI/F770W flux even if we do not include that observation in our {\tt emcee} fit.

\begin{figure}
    \begin{center}
    \includegraphics[width=0.4\textwidth]{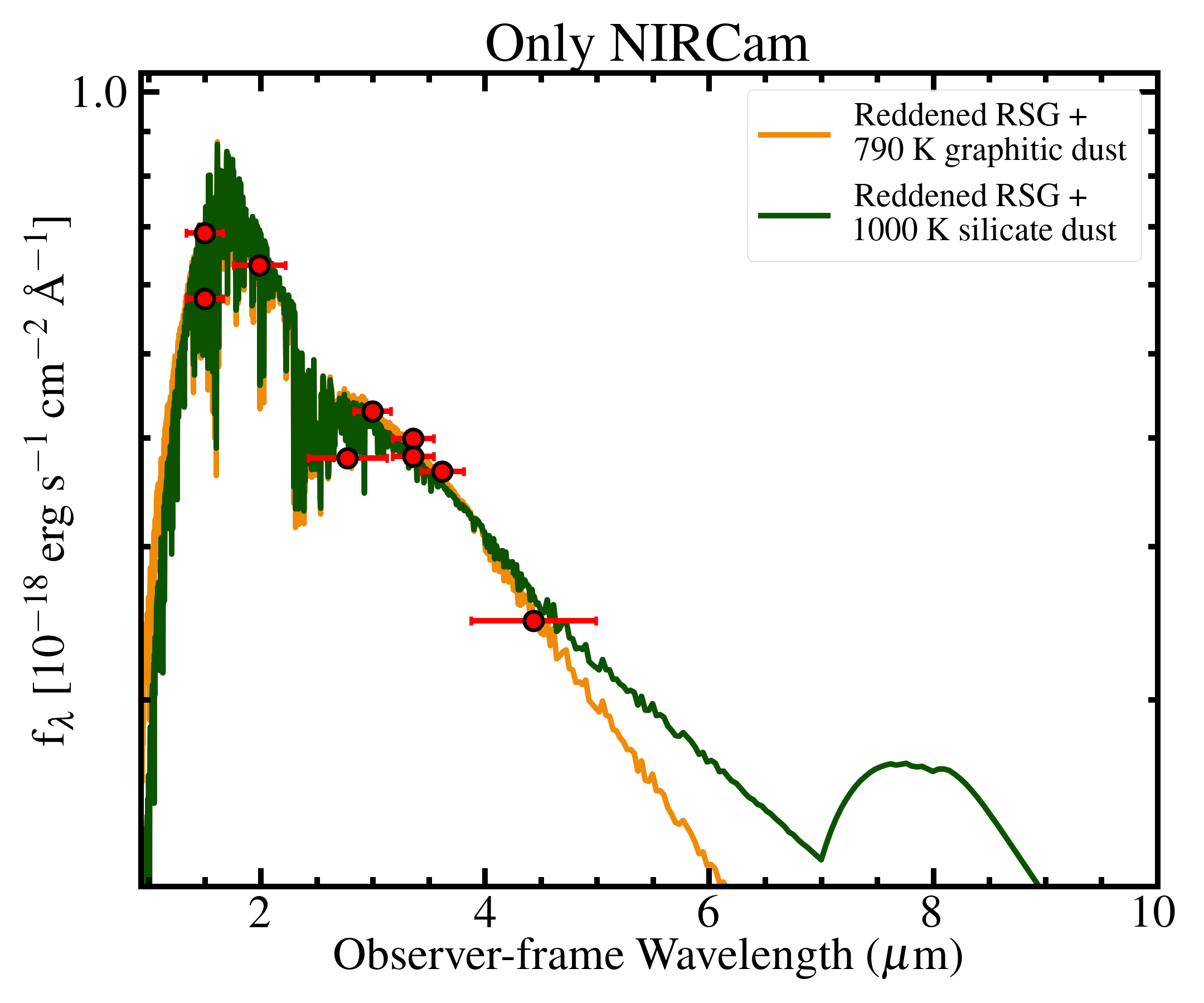}
    \vspace{-0.5in}
    \includegraphics[width=0.4\textwidth]{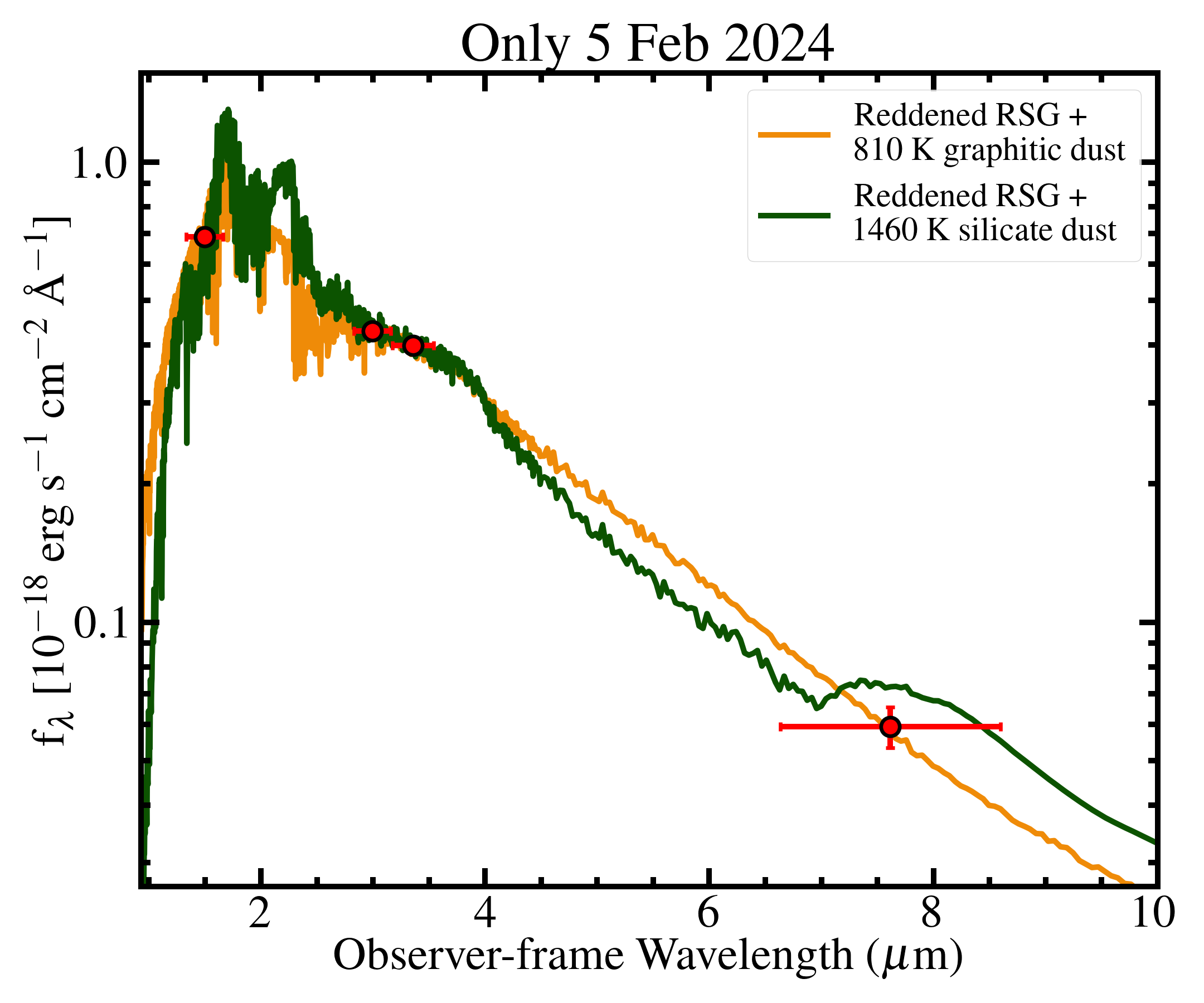}
    \vspace{-0.5in}
    \includegraphics[width=0.4\textwidth]{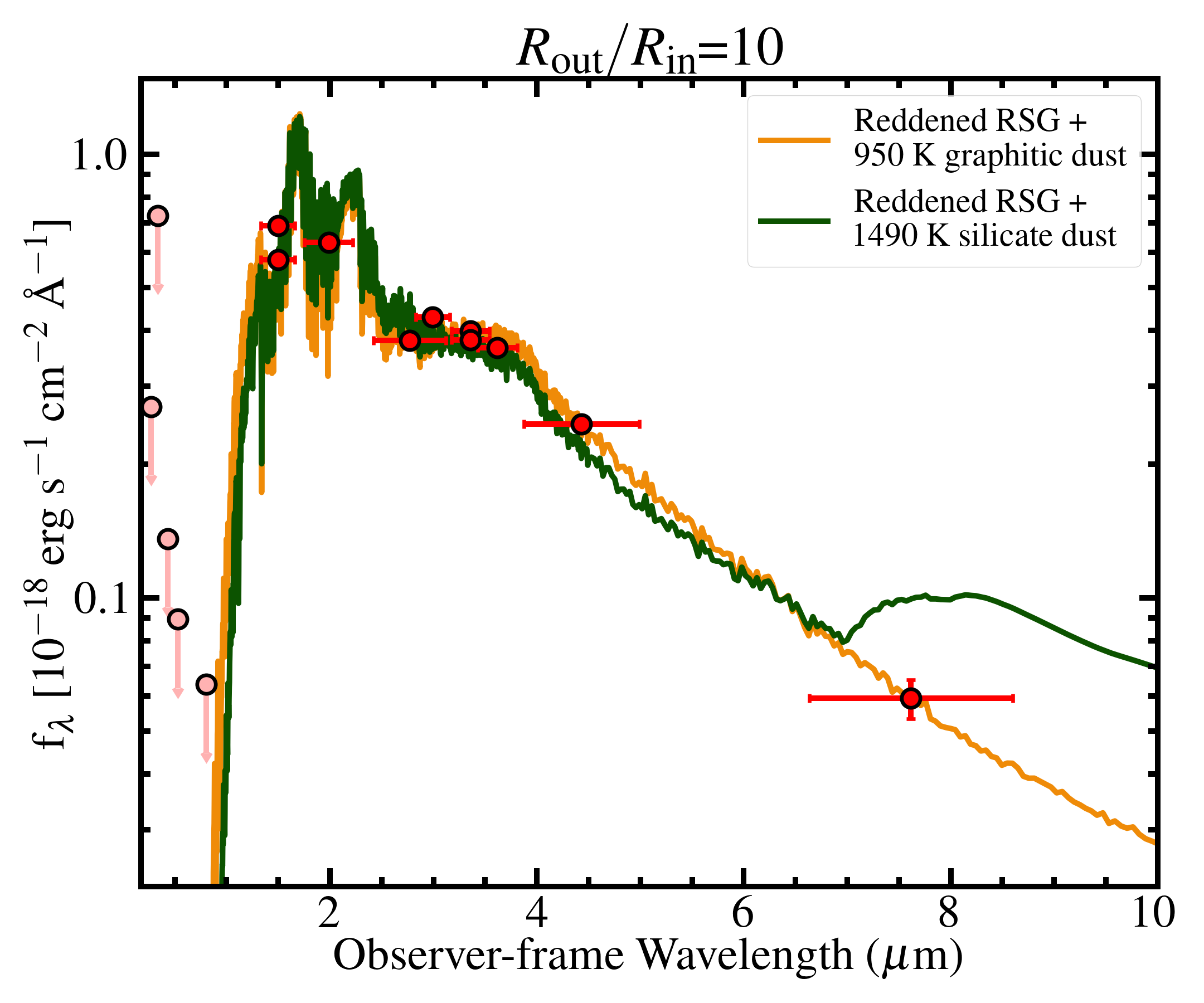}
    \end{center}
    \caption{{\it Top}: Same as Figure~\ref{fig:sed}, but for graphitic and silicate dust models fit only to the {\it JWST}/NIRCam photometry.  This model results in similar posteriors to the full SED fit.  {\it Middle}: Same as the top, but using only {\it JWST}/NIRCam and MIRI photometry from 5 Feb 2024 to minimize the effects of variability.  The posteriors for the silicate model still overpredict the F770W flux.  {\it Bottom}: Same as Figure~\ref{fig:sed}, but setting the dust shell boundaries using $R_{\rm out}/R_{\rm in}=10$ in our model as described in Section~\ref{sec:rsg_sed}.  The silicate model again overpredicts the F770W flux.}\label{fig:models}
\end{figure}

We next consider the possibility that, although the $\sigma_{\rm sys}$ value derived from the silicate-rich fit is comparable to the level of observed variability in F150W and F335M in Section~\ref{sec:variability}, the variability in F770W could be larger and thus combining photometry over a $\approx$8~month period is not representative of the underlying SED \citep[similar to SN\,2023ixf;][]{Kilpatrick23,Jencson23}.  We perform a new fit by restricting the model to constrain only wide-band detections obtained on 5 Feb 2024 (Figure~\ref{fig:models}), which were observed within the span of $\approx$2~hr and are therefore unlikely to be significantly affected by variability in any RSG \citep[whose characteristic variability timescales are months to years;][]{Soraisam18}.  Again, the model predicts a large $\tau_{V}=22.7\substack{+0.6\\-0.8}$ with a much higher flux in F770W than we observe.

Finally, we consider the possibility that our choice of the dust shell width has some effect on the observed level of dust opacity and associated emission in the F770W band.  We try a new fit with a moderately extended dust shell where $R_{\rm out}=10 R_{\rm in}$.  The overall opacity prediction remains high at $\tau_{V}=26.1$, and the model remains inconsistent with the F770W detection at $>$3$\sigma$ whereas the graphitic model is consistent with the data for this dust shell width.  Moreover, the dust shell cannot be significantly wider than this as the radius predicted by this silicate model is $R_{\rm in}=5240~R_{\odot}$, implying that the outer radius $52,400~R_{\odot}$ is approaching our limit on the size of the dust shell from spectroscopy.

We therefore conclude that the SN\,2025pht counterpart photometry is consistent with a RSG enshrouded in graphitic dust shell and inconsistent with silicate-rich circumstellar dust.  The underlying RSG is robustly predicted as having $\log(L/L_{\odot})=5.0$ due to the excellent sampling of the SED, even within relatively narrow time windows.  Our best-fitting graphitic model predicts a moderately high optical opacity of $\tau_{V}=6.7$, which corresponds to $A_{V}=5.3$~mag of circumstellar extinction due to the dust shell.  We discuss these conclusions and the implications for RSG modeling and SN progenitors below.

\section{Discussion}\label{sec:discussion}

The SN\,2025pht progenitor candidate is on the high-luminosity end of RSG counterparts to nearby SN\,II observed to date, although not the most luminous system observed \citep[which is likely SN\,2009hd at $\log(L/L_{\odot})=5.24$;][]{Healy24} or beyond the ``maximum luminosity'' predicted in analyses of the RSG problem \citep[e.g., in][]{Smartt09,Kochanek20,Beasor+2025}.  Nevertheless, compared with the vast majority of SN\,II pre-explosion counterparts, which are often detected in only a single band, its luminosity is robustly constrained via detections from WFPC2/F814W to MIRI/F770W, roughly an order of magnitude in wavelength.  In particular, constraints on the magnitude and nature of dust emission in the circumstellar environment around this source are more tightly constrained than almost any other SN\,II progenitor candidate \citep[e.g., in][]{Smartt15,Kilpatrick18:17eaw,vandyk+18,Kilpatrick23,Jencson23}.

\subsection{Carbon-rich Dust in the Circumstellar Environment of SN~2025pht}

SN\,2025pht is consistent with a $\log(L/L_{\odot})=5.0$ RSG, which at the metallicity of its host galaxy would imply a single star with an initial mass $M_{\rm ZAMS}=15~M_{\odot}$ \citep{choi+16}.  In general, RSGs in this mass range and above exhibit silicate-rich dust profiles based on spectroscopy \citep{verhoelst+09}.  However, photometric fitting for large populations of RSGs in M31 and M33 that include near-IR and WISE mid-IR photometry indicate that carbon-rich dust is rarer but still observed for some RSGs in this luminosity range \citep{Wang21}.  Overall, these RSGs span a relatively narrow range in temperature and metallicity, and so we infer that differences in the composition of their circumstellar dust is largely driven by their surface abundances \citep[as in][]{Lancon07} as opposed to their environments or the relative fragility of dust grains to the star's spectrum.

In general, carbon-rich dust molecules require a circumstellar C/O ratio greater than 1 such that carbon atoms are left over after the formation of carbon monoxide \citep{Savage1979}.  \citet{Davies19} argue that as we look at higher $M_{\rm init}$ stars, their larger convective cores will in turn yield a larger fraction of CNO-processed material in the envelope of the star at the onset of core He burning and the RSG phase.  This implies that the mass fraction of carbon will be depleted, and thus we expect monotonically decreasing C/N and C/O ratios with respect to $M_{\rm init}$, which would yield a higher fraction of stars with silicate-rich circumstellar environments at the point of core collapse.  This effect can be mitigated somewhat by assumptions of rotation and convective overshoot \citep{Maeder00,Ekstrom12}, and so we emphasize that the final C/O ratio is uncertain by nearly 1~dex even in \citet{Davies19}.  Moreover, the fact that some RSGs in this luminosity range exhibit graphitic dust is evidence that their surface carbon abundances can be large.

On the other hand, we note that flash spectra of SN\,II favor stars with surface abundances C/N$<$1 and C/O$<$1 \citep[see][]{Dessart23,Jacobson-Galan23,Jacobson-Galan24,Dessart25}, in agreement with the values predicted by \citet{Davies19}.  In light of these data, a high fraction of terminal RSGs with enhanced C/N and C/O ratios in the $\approx$12--15~$M_{\odot}$ mass range would be intriguing, especially as they are not expected to undergo the post He-burning dredge up that produces carbon stars in lower mass ranges \citep[e.g., for stars $<$9.2~$M_{\odot}$;][]{Sugimoto1980,Limongi24}.  We note, however, that the presence of dense CSM confined closely around a large fraction of SN\,II \citep[based on light curve modeling and flash spectroscopy;][]{khazov+15,Forster18,Morozova18,Bruch21} suggests an energetic mechanism that can release a large fraction of the star's mass \citep[e.g., a ``superwind'' or outburst as in][]{Morozova20,Davies22,Fuller24}.  If this mechanism is energetic enough to initiate post He-burning dredge up from the base of the RSG's convective envelope, it could simultaneously explain both the presence and composition of the CSM around SN progenitor stars similar to SN\,2025pht.  Future studies that analyze the correlation between the mass and composition of this material may offer a way to determine its exact origin in the progenitor system.

\subsection{The High Optical Opacity Toward the SN~2025\lowercase{pht} Counterpart and Implications for the Red Supergiant Problem}\label{sec:bc}

As discussed above, the vast majority of SN\,II pre-explosion counterparts are observed in F814W as the reddest (and sometimes only) band in their SEDs \citep[see joint analyses in][]{Smartt09,Smartt15,Davies18,Kochanek20,Healy24}.  This places a significant systematic uncertainty on their bolometric corrections when taking into consideration that their optical opacities can be as high as that of SN\,2025pht.  We underscore this point here by comparing bolometric corrections from the recent joint analysis in \citet{Healy24} to the values we infer from well-sampled SN progenitor star SEDs.

\citet{Healy24} derive F814W, $I_{c}$- (Cousins), $I_{j}$- (Johnson), or $i$- (SDSS) band bolometric corrections for 22/30 SN pre-explosion counterparts including uncertainties.  We treat these bands as being comparable in wavelength and model their probability density functions (PDFs) as Gaussian distributions with the given means and standard deviations.  We show the joint PDF for 20 of these sources discounting SN\,2023ixf and 2024ggi in Figure~\ref{fig:bcs}.  For comparison, we show the bolometric corrections for SN\,2017eaw \citep{Kilpatrick18:17eaw}, 2023ixf \citep{Kilpatrick23}, 2024ggi \citep{Xiang24}, and 2025pht (this paper), which were selected since their counterparts have $>$2~$\mu$m detections from {\it Spitzer}/IRAC or {\it JWST}.  We model these bolometric corrections by running the same reddened RSG model described in Section~\ref{sec:rsg_sed} for MARCS SEDs surrounded by a graphitic, $R_{\rm out}/R_{\rm in}=2$ dust shell (see the derived model parameters in Table~\ref{tab:bc_models}).  Each bolometric correction we derive is modeled to the WFC3/UVIS F814W band by taking 100,000 samples from the model posteriors and calculating $M_{\rm bol} - M_{\rm F814W}$, which we show in Figure~\ref{fig:bcs}.  We emphasize that since the four objects we analyze all have well-constrained SEDs from optical to mid-IR wavelengths, the bolometric corrections are relatively insensitive to the choice of dust model and our conclusions below depend solely on the derived luminosities.

\begin{deluxetable*}{ccccccc}
\tablecaption{Derived Parameters for the SN\,2025pht Progenitor Star and Other Progenitor Stars from the Literature}
\tablehead{
\colhead{SN Name} &
\colhead{$\log(L/L_{\odot})$} &
\colhead{$T_{\rm eff}$} &
\colhead{$T_{\rm dust}$} &
\colhead{$\tau_{V}$} &
\colhead{$\sigma_{\rm sys}$} &
\colhead{References} \\
&
&
(K) &
(K) &
&
(mag) &
}
\startdata
{\bf 2025pht} & 5.00$\substack{+0.09\\-0.08}$ & 3030$\substack{+210\\-270}$ & 790$\substack{+40\\-30}$ & 6.7$\pm$0.4 & 0.08$\substack{+0.03\\-0.02}$ & (this paper) \\
2017eaw & 4.90$\pm$0.08 & 3310$\substack{+40\\-30}$ & 910$\substack{+300\\-280}$ & 1.2$\pm$0.4 & 0.13$\substack{+0.06\\-0.04}$ & [1] \\
2023ixf & 4.73$\substack{+0.10\\-0.08}$ & 4000$\substack{+330\\-290}$ & 1180$\substack{+130\\-120}$ & 5.8$\pm$0.3 & 0.09$\substack{+0.10\\-0.06}$ & [2] \\
2024ggi & 4.85$\pm$0.12 & 3280$\substack{+70\\-110}$ & 660$\substack{+200\\-120}$ & 1.6$\substack{+0.4\\-0.5}$ & 0.32$\substack{+0.05\\-0.04}$ & [3]
\enddata
\tablecomments{The derived model parameters from running our graphitic, $R_{\rm out}/R_{\rm in}=2$ RSG dust model as described in Section~\ref{sec:rsg_sed} for SN\,2025pht and several SN progenitors with photometry presented in the literature.  The photometry comes from the provided references: [1] \citet{Kilpatrick18:17eaw}, [2] \citet{Kilpatrick23}, [3] \citet{Xiang24}.}\label{tab:bc_models}
\end{deluxetable*}

\begin{figure}
    \includegraphics[width=0.49\textwidth]{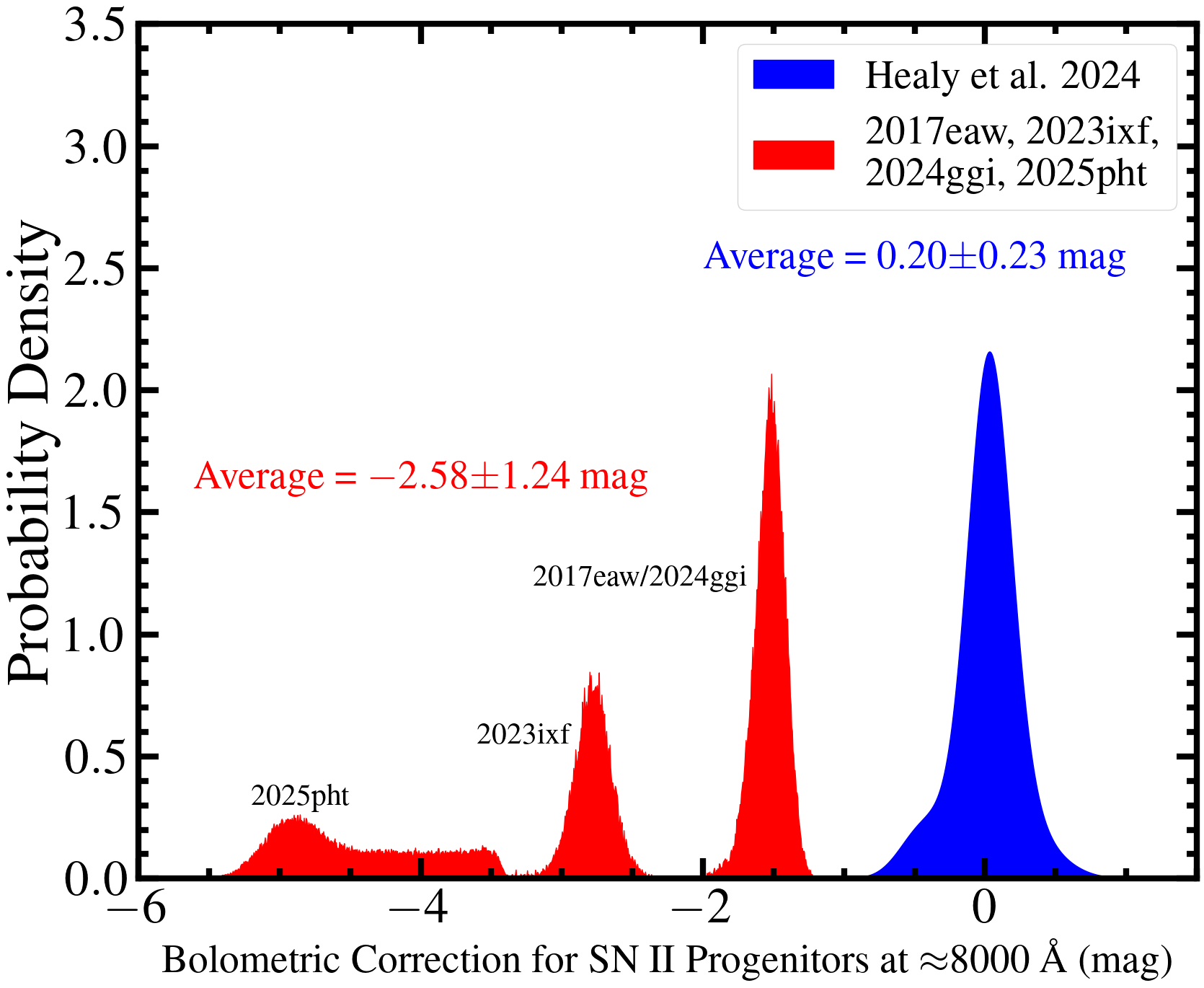}
    \caption{The probability density for the bolometric correction to $\approx$8000~\AA\ (i.e., for F814W, $I_{c}$, $I_{j}$, and $i$-bands) for 20 pre-SN counterparts to Type\,II SNe from \citet{Healy24} (blue; see their Table~A.2) compared with the bolometric correction derived for WFC3/UVIS F814W in our modeling of SN\,2017eaw, 2023ixf, 2024ggi, and 2025pht (red; see Section~\ref{sec:bc}).  As these latter events have well-sampled SEDs, including at $>$2~$\mu$m with {\it Spitzer} and {\it JWST}, we consider their optical bolometric corrections where most SN\,II counterparts are detected to be more reliable than for SN\,II counterparts that lack IR photometry.  We measure the average and standard deviation of both distributions as 0.20$\pm$0.23~mag and -2.58$\pm$1.24~mag.}\label{fig:bcs}
\end{figure}

We caution that the four objects we analyze may not be representative of the overall SN\,II population due to selection effects associated with requiring $>$2~$\mu$m detections \citep[although many other sources have constraining upper limits at these wavelengths, e.g., for SN\,2020jfo;][]{Kilpatrick23:20jfo} or from possible biases in the photometry used, our modeling, or the more systematic modeling in studies such as \citet{Healy24}.  However, we note that these four sources all have systematically larger bolometric corrections than those analyzed in the literature.  We infer that a significant fraction of the sources with only F814W detections are actually much more luminous than assumed \citep[similar to conclusions in][]{Beasor+2025}.  If we consider the difference between the average of the distribution of bolometric corrections we calculate and those from \citet{Healy24}, which is 2.78$\pm$1.26~mag or 1.1$\pm$0.50~dex in luminosity, a correction even on the lower side of this distribution could place low-luminosity SN\,II counterparts \citep[such as SN\,2003gd or 2020jfo with $\log(L/L_{\odot})=4.3$ and $\log(L/L_{\odot})=4.1$, respectively;][]{Eldridge+04,Sollerman21,Teja22,Kilpatrick23:20jfo} above the ``maximum'' luminosity counterpart of $\log(L/L_{\odot})=5.2$.  Thus we emphasize that more detailed modeling of all such RSG counterparts is warranted in light of the strong priors on high circumstellar extinction from more recent SN\,II progenitor star detections.

\subsection{Caveats to our Inferences on the Nature of the SN~2025pht Progenitor and Future Direction}

The conclusions listed above for SN\,2025pht assume a fixed dust abundance profile with only two opacity laws from \citet{Draine1984}, a spherical shell following a $r^{-2}$ wind profile within fixed radii, an arbitrary dust grain size distribution, and we largely ignore the effects of variability in the progenitor.  We have attempted to quantify some of these effects, and in particular we investigated the effects of variability in Section~\ref{sec:rsg_sed} and found no significant change in the preferred dust model.  Moreover, we do not expect these effects to change the overall luminosity of the system given the sampling of the SED or the finding that the SN\,2025pht progenitor system experienced a large magnitude of circumstellar extinction. 

Nevertheless, future analyses can account for some of these uncertainties in dust properties within the constraints of physically-motivated priors by fitting RSG SEDs with: i) a variable dust composition between pure graphitic, pure silicate, or some arbitrary mixture of the two, ii) variable dust geometries following a range of radial profiles and inner and outer dust shell ratios, and iii) time-variable fits restricted to narrow ranges in time as in Figure~\ref{fig:models}.  While it is possible that, via some combination of the above, we could find a range of parameter space that favor dust compositions other than graphitic, our best-fitting models consistently favor the highly-extinguished $\log(L/L_{\odot})=5.0$ graphitic dust model discussed above.

Existing and future {\it JWST} observations \citep[as well as {\it Nancy Grace Roman Space Telescope} imaging;][]{Gezari22,Lancaster22,Dage23} will enable these more sophisticated dust fitting prescriptions even for the vast population of RSGs observed in situ for nearby galaxies such as NGC\,1637.  Indeed, the peculiar nature of the SN\,2025pht progenitor, and in particular its highly-reddened SED consistent with graphitic dust, may be an avenue for identifying RSGs in the final phases of evolution before explosion \citep[i.e., ``superwind'' or post-outburst RSGs;][]{Morozova20,Davies22}.  Along with follow up of low-luminosity eruptive sources coincident with RSGs \citep[similar to SN\,2020tlf;][]{Jacobson-Galan22} and other highly-variable RSGs \citep[similar to SN\,2023ixf;][]{Kilpatrick23,Jencson23}, surveying these mid-IR bright RSGs may be the most promising avenue to obtaining spectral classification of a SN progenitor star in the final months of years before explosion.  This would allow us to definitively answer questions about spectral type, stellar mass, metallicity, circumstellar extinction, dust composition, and the presence of a companion star in SN progenitor systems that are obscured via broadband SED fitting.

\section{Conclusions}

We present an analysis of the Type\,II SN 2025pht in NGC\,1637 at 12~Mpc and the detection of its counterpart in pre-explosion {\it JWST} and {\it HST} imaging.  From our analysis, we find:

\begin{enumerate}
    \item The counterpart is detected at wavelengths from WFPC2/F814W ($\approx$0.8~$\mu$m) to MIRI/F770W (7.6~$\mu$m) representing one of the best-sampled SN progenitor candidates to date in wavelength, including the first detection of a counterpart at wavelengths $>$5~$\mu$m.
    \item From these data, we fit dust-obscured RSG SEDs and find that the source is moderately luminous with $\log(L/L_{\odot})=5.0$ (corresponding to an initial mass of $\approx$15~$M_{\odot}$) but also heavily dust-obscured.  In our best-fitting model, we prefer graphite-rich dust with a high line-of-sight optical extinction of $A_{V}=5.3$~mag.
    \item The implied carbon-rich dust composition is unusual for high-mass RSGs that tend to be oxygen-rich and produce silicates in their circumstellar environments.  If the SN\,2025pht counterpart had a carbon-rich circumstellar environment, it may imply that the surface abundances of some evolved, carbon-burning and later RSGs are significantly enhanced due to convection.
    \item Along with SN\,2017eaw, 2023ixf, and 2024ggi, SN\,2025pht belongs to a class of sources with pre-explosion counterparts detected at $>$2~$\mu$m.  These sources are all consistent with moderate to heavy circumstellar reddening in the underlying RSG, significantly in excess of previous predictions on the magnitude of circumstellar extinction for SN progenitor stars.  New analyses of the overall population of SN\,II progenitor stars that incorporate these constraints on their typical circumstellar extinction may mitigate or entirely erase the mismatch between their luminosities and the overall populations of RSGs.
\end{enumerate}

SN\,2025pht represents the beginning of SN progenitor star analyses performed with {\it JWST}.  As shown above, the broad wavelength coverage and precise photometry enabled by NIRCam and MIRI can enable novel constraints on the nature of the circumstellar environment around these sources and more precise estimates of their bolometric luminosities.  By answering long-standing questions about the terminal states of massive stars, we are now better able to bridge the gap between direct imaging of SN progenitor systems and indirect constraints from their pre-explosion outbursts \citep{Jacobson-Galan22}, early light curves \citep{Hosseinzadeh18,Hosseinzadeh22,Kilpatrick23:20jfo}, and flash spectroscopy \citep{Jacobson-Galan24}.

\bigskip\bigskip\bigskip
\noindent {\bf ACKNOWLEDGMENTS}
\smallskip

We would like to thank STScI scientists Amber Armstrong, Joel Green, and Christian Soto for supporting and triggering HST-GO-17706.

C.D.K. and A.S. gratefully acknowledge support from the NSF through AST-2432037, the HST Guest Observer Program through HST-SNAP-17070 and HST-GO-17706, and from JWST Archival Research through JWST-AR-6241 and JWST-AR-5441.
The UCSC team is supported in part by NASA grants 80NSSC23K0301 and 80NSSC24K1411; JWST Archival Research through JWST-AR-6241 and JWST-AR-5441; and a fellowship from the David and Lucile Packard Foundation to R.J.F.
M.R.D acknowledges support from the NSERC through grant RGPIN-2019-06186, the Canada Research Chairs Program, and the Dunlap Institute at the University of Toronto.
A.~G. is supported by the National Science Foundation under Cooperative Agreement PHY-2019786 (The NSF AI Institute for Artificial Intelligence and Fundamental Interactions, http://iaifi.org/).
W.J.-G. is supported by NASA through Hubble Fellowship grant HSTHF2-51558.001-A awarded by the Space Telescope Science Institute, which is operated for NASA by the Association of Universities for Research in Astronomy, Inc., under contract NAS5-26555.

This research is based on observations made with the NASA/ESA Hubble Space Telescope obtained from the Space Telescope Science Institute, which is operated by the Association of Universities for Research in Astronomy, Inc., under NASA contract NAS 5–26555. These observations are associated with programs HST-GO-5446 (PI Illingworth), HST-GO-9041 (PI Smartt), HST-SNAP-9042 (PI Smartt), HST-GO-9155 (PI Leonard), HST-GO-17502 (PI Thilker), and HST-GO-17706 (PI Kilpatrick).

This work is based in part on observations made with the NASA/ESA/CSA James Webb Space Telescope. The data were obtained from the Mikulski Archive for Space Telescopes at the Space Telescope Science Institute, which is operated by the Association of Universities for Research in Astronomy, Inc., under NASA contract NAS 5-03127 for JWST. These observations are associated with programs JWST-GO-3707 (PI Leroy) and JWST-GO-4793 (PI Shinnerer).

Some of the data presented herein were obtained at Keck Observatory, which is a private 501(c)3 non-profit organization operated as a scientific partnership among the California Institute of Technology, the University of California, and the National Aeronautics and Space Administration. The Observatory was made possible by the generous financial support of the W.\ M.\ Keck Foundation.

The authors wish to recognize and acknowledge the very significant cultural role and reverence that the summit of Maunakea has always had within the Native Hawaiian community. We are most fortunate to have the opportunity to conduct observations from this mountain.

\textit{Facilities}: {\it HST} (WFPC2, WFC3), {\it JWST} (NIRCam, MIRI), Keck:I (LRIS)

\textit{Software}:
{\tt astropy} \citep{astropy},
{\tt dolphot} \citep{dolphot},
{\tt drizzlepac} \citep{drizzlepac},
{\tt dusty} \citep{dusty},
{\tt emcee} \citep{emcee},
{\tt extinction} \citep{extinction},
{\tt hst123} \citep{hst123},
{\tt JHAT} \citep{jhat},
{\tt SNID} \citep{snid},
{\tt synphot} \citep{synphot},
{\tt TweakReg} \citep{tweakreg},
{\tt UCSC Spectral Pipeline} \citep{Siebert20},
{\tt progenitors} \citep{progenitors}

\bigskip

\section*{Data and Software Availability}

All data and analysis products presented in this article are available upon request.  Analysis code and photometry used in this paper are available at \url{https://github.com/charliekilpatrick/progenitors}.  
The {\it Hubble Space Telescope} and {\it James Webb Space Telescope} data used in this paper can be found in MAST: \dataset[10.17909/dj00-e351]{http://dx.doi.org/10.17909/dj00-e351}.  

% REFERENCES
\bibliographystyle{aasjournal} 
\bibliography{2025pht}

\end{document}